\documentclass{article}
\usepackage{INTERSPEECH2020,amsmath,graphicx}
\usepackage[font=small,skip=0pt]{caption}
\usepackage{amsfonts}
\usepackage{wrapfig}
\usepackage{amsmath,bm}

\title{Speaker Independent and Multilingual/Mixlingual Speech-Driven Talking Head Generation Using Phonetic Posteriorgrams}

 \name{\begin{tabular}{c}Huirong Huang$^{1, 2, 3}$, Zhiyong Wu$^{1,3,4}$, Shiyin Kang$^{2,*}$\thanks{* Corresponding author} \thanks{This work is done during the internship of first author at Tencent.}, Dongyang Dai$^{1,3}$, Jia Jia$^{1,3}$\\ Tianxiao Fu$^{2}$, Deyi Tuo$^{2}$, Guangzhi Lei$^{2}$, Peng Liu$^{2}$, Dan Su$^{2}$, Dong Yu$^{2}$, Helen Meng$^{1,4}$\end{tabular}}

 \address{ $^1$Tsinghua-CUHK Joint Research Center for Media Sciences, Technologies and Systems, \\
 	 Shenzhen International Graduate School, Tsinghua University, Shenzhen, China\\
 	 $^2$Tencent AI LAB, Tencent, Shenzhen, China\\
 	 $^3$Department of Computer Science and Technology, Tsinghua University, Beijing, China \\
 	 $^4$Department of Systems Engineering and Engineering Management, \\
 	 The Chinese University of Hong Kong, Shatin, N.T., Hong Kong SAR, China
 	}
 \email{hhr18@mails.tsinghua.edu.cn, \{zywu,hmmeng\}@se.cuhk.edu.hk, daizhouyin@163.com,\\
 \{shiyinkang,alvinfu,deyituo,guangzhilei,feanorliu,dansu, dyu\}@tencent.com}
\begin{document}

%
\maketitle
\begin{abstract}
Generating 3D speech-driven talking head has received more and more attention in recent years. Recent approaches mainly have following limitations: 1) most speaker-independent methods need handcrafted features that are time-consuming to design or unreliable; 2) there is no convincing method to support multilingual or mixlingual speech as input. In this work, we propose a novel approach using phonetic posteriorgrams (PPG). In this way, our method doesn't need hand-crafted features and is more robust to noise compared to recent approaches. Furthermore, our method can support multilingual speech as input by building a universal phoneme space. As far as we know, our model is the first to support multilingual/mixlingual speech as input with convincing results. Objective and subjective experiments have shown that our model can generate high quality animations given speech from unseen languages or speakers and be robust to noise.
\end{abstract}
\noindent\textbf{Index Terms}: 
speaker independency, multilingual generation, phonetic posteriorgrams (PPG), universal phoneme space, talking head
\section{Introduction}

With the development of machine learning and 3D animation technology, 3D virtual talking head has become an important research topic in recent years. To be specific, 3D talking head animation is to generate expressions and lip movements synchronized with the input speech by modifying the facial parameters of the 3D talking head. One of the most important applications is speech driven facial animation generation which accepts speech as input and generates corresponding facial animations with 3D talking head. Conventional speech driven animation methods applied in 3D game or movie production often require a professional animation production team which is time-consuming and expensive.
 
To solve the problems of conventional methods, data-driven automatic facial animation generation methods have received a lot of attention recently. There mainly exist three data-driven methods to generate animations: unit-selection based method, Hidden Markov Model (HMM) based method and machine learning based method. In these methods, how to model the contextual information is crucial for generating accurate facial animation. Unit-selection based method is able to generate natural facial animations, but it needs a large amount of data to cover different phonetic context cases \cite{xu2013practical}\cite{taylor2012dynamic}\cite{theobald2012relating}\cite{mattheyses2013comprehensive}. HMM based method achieves poor performance because the limited parameters are not enough to capture coarticulation effects \cite{merritt2013investigating}\cite{wang2012high}\cite{schabus2011simultaneous}. Compared to these two methods, machine learning method like recurrent neural network (RNN) or convolution neural network (CNN) can capture coarticulation effects and be robust to unseen patterns \cite{obamanet}\cite{obamalip}\cite{dahmani2019conditional}\cite{fehp2018}\cite{biasutto2019modeling}. Based on the above reasons, we apply bidirectional long short-term memory (BLSTM) model in this paper which shows its ability to accurately predict facial animation parameters.

Methods mentioned above mainly apply speaker-dependent approach to generate animations because multi-speaker speech-to-face datasets are difficult to obtain. Since we obviously cannot collect data from all users for training in real applications, it is desirable to develop speaker-independent method to generate animations from unseen speakers. In dealing with the problem, recent works have focused on generating animations through single-speaker datasets by using phoneme sequences or pre-trained methods. In \cite{disney}, phoneme sequences are used as input features in order to remove speaker related information. But phoneme sequences destroy the information of phoneme distribution and duration causing poor prediction. Some hand-crafted features have to be added to improve the performance \cite{disney}. A similar work \cite{xielei2016} uses phoneme sequences as input features as well. Different from \cite{disney}, \cite{xielei2016} adds phoneme state of each frame to improve the distinguishability. All the works need hand-crafted features that take much time cost to design and are not able to keep content information well. \cite{yousay} achieves satisfactory results using pre-trained CNN architecture that is similar to VGG-M \cite{vgg}, but input features of this method are speaker-dependent.

Multilingual/mixlingual generation is also an especially disired feature in many scenarios. For example in 3D animation creation, multilingual/mixlingual issues arise when the voice actors come from multiple countries. Recently proposed models are language-dependent of which most only support monolingual speech. The only related work we find is \cite{disney} but no convincing experimental result is presented. In other words, there is currently no model that can support multilingual/mixlingual generation.

To achieve speaker-independency, we need to explore a robust method that doesn't use speaker-dependent features like F0 or mel-frequency cepstral coefficients (MFCC). Some recent voice conversion works like \cite{ppg} and \cite{luhui} apply PPG to remove speaker-dependent information and achieve a good performance. Compared to phoneme sequences, PPG keeps the phoneme distributions which contains durations implicitly. So it can be used to obtain more accurate predictions while keeping speaker-independent. Inspired by above works, we apply PPG to our model and our experimental results show significant improvement over speaker-dependent models. Compared to recent models, our model doesn't require hand-crafted features and multi-speaker speech-to-face datasets while achieving better results in multi-speaker cases.

Besides, we propose a novel language-independent method by introducing universal phoneme space. The phoneme set of PPG can be extended to phonemes of multiple languages so that PPG can be trained with multilingual ASR dataset. In this way, there is no need for multilingual speech-to-face dataset and new language can be easily appended to PPG. Using the trained multilingual PPG, our model can generate high quality multilingual/mixlingual animations as our experimental results present.

\section{Proposed Method}

\begin{figure}[t]
  \centering
  \includegraphics[width=5.8cm, height=7cm]{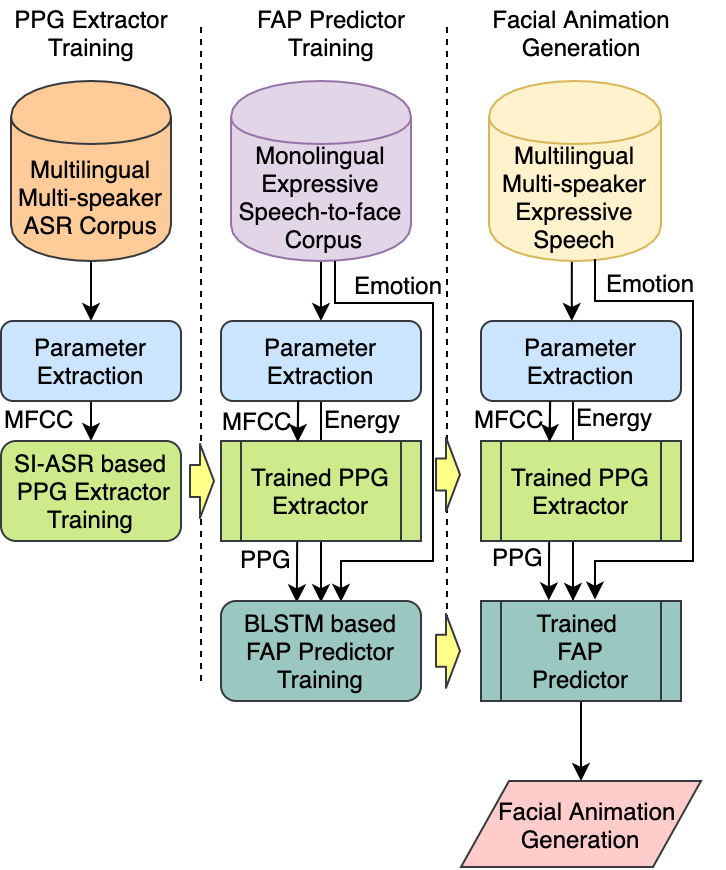}
  \vspace{8pt}
  \caption{Overview of the proposed method}
  \label{fig:system}
  \vspace{-13pt}
\end{figure}

Our model mainly consists of 3 modules: PPG extraction module, facial animation parameters (FAP) prediction module and facial animation generation module. The pipeline is: 1) Train speaker independent automatic speech recognition (SI-ASR) based PPG extractor using multilingual/mixlingual ASR corpus; 2) Extract PPG features from trained PPG extractor and train BLSTM based FAP predictor with monolingual speech-to-face corpus; 3) Generate facial animation parameters from any speaker's speech using trained PPG extractor and FAP predictor. Fig.\ref{fig:system} depicts a detailed overview of our proposed method.

\subsection{PPG Extraction}
\label{ssec:ASR module part}
Recent speaker-independent facial animation generation methods use phoneme sequences as input feature. Even though phoneme sequence is speaker-independent, it can't retain phoneme distribution and duration information that are important for facial animation parameters prediction. In this work, we propose a method using PPG. PPG is a time sequence representing the poterior probabilities of each phonetic unit for every time step. Trained on noisy multi-speaker dataset, PPG can equalize speaker differences. While keeping speaker-independent, PPG also retains phoneme distribution and duration information which can be revealed by number of frames each phonetic unit lasts.

In this paper, we extract PPG using a SI-ASR model with multi-speaker ASR dataset. The basic unit in PPG's phoneme space is monophone so that our dataset can cover most phoneme units. In terms of model settings, the SI-ASR model is built with 1 convolutional-1D layer and 4 dense layers with 512 units. For input features, 40-dimension filter-bank features are used. Since context has a great influence on recognition results, we add a context of $\pm$10 frames, first order difference and second order difference to the input.

The problem of multilingualism in speech-to-face is very challenging since multilingual speech-to-face datasets are difficult to obtain. However, multilingual ASR dataset is easily available. If we could incorporate the multilingual information into PPGs, prediction of facial animation parameters might possess the generalization ability to multiple languages. In this way, multilingual/mixlingual speech driven facial animations can be achieved. This motivates us to propose a method based on universal phoneme space. Universal phoneme space means that the PPG phoneme set contains phonemes of multiple languages. In this way, each group of phonemes in PPG corresponding to specific language can be learned independently using a multilingual ASR dataset. After that, multilingual PPG makes the model robust to multilingual input. Besides, universal phoneme space makes the model more scalable. To support a new language, the model only need to append new phoneme units and train the PPG using ASR dataset of the new language.

\subsection{Facial animation parameters prediction}
\label{ssec:conversion part}

To predict facial animation parameters from PPG, we build a conversion network using BLSTM recurrent neural network. Due to the effect of coarticulation, speech sound is influenced by its neighboring sounds. For this reason, BLSTM is used to model the context information. In terms of architecture, our model consists of three 128-unit BLSTM layers and two 96-unit dense layers. As for output features, facial animation parameters are 32 dimensional face warehouse parameters \cite{fwh32} which are extracted from 3D blendshape of real person. To be specific, the first 25 dimensions represent facial animation parameters of eyes and mouth. The other 7 dimensions mainly indicate 3D head pose. During training, zoneout \cite{zone} is introduced to the BLSTM layers to prevent the model from overfitting. According to the experimental results, our model achieves the best results when zoneout rate is set to 0.1.

Besides, we have also tried a sliding window CNN model inspired by \cite{disney} to predict facial animation parameters. We set a similar encoder like \cite{yousay} except that, instead of using 2D convolution, we use 1D convolution network in our case because different dimensions of PPG features (corresponding to different phonetic units) are independent. In generation stage, we generate facial animation parameter sequences using sliding window and then average frame-wise. Experiments have shown that CNN performs sligtly worse than BLSTM model which mainly results from the discontinuity between adjacent sliding windows. Therefore, in the following experiments, we all use BLSTM as the conversion network for reasonable comparison of different models.

However, generated animation with PPG as input only has a problem that the mouth cannot keep closed during silent part in speech. To deal with the problem, a method is disired to distinguish the silent part from speech. Since silent part has much lower energy, the frame level energy of the input speech is introduced as the additional input of BLSTM based conversion network for better prediction of mouth related parameters.

In order to make the generated animation more expressive, we built an expressive speech-to-face dataset with multiple emotions. With this dataset, an expressive animation parameters predictor can be realized by augmenting the aforementioned BLSTM model with additional emotion label input. In this way, our model can learn different animation patterns of different emotions instead of an average pattern of all emotions. At the stage of generation, expressive animations can be easily generated by feeding different emotion labels.

\subsection{Facial animation generation}
\label{ssec:postprocess generation part}

Generating animations from predicted face warehouse parameters directly will cause discontinuity problems. To alleviate the problem, maximum likelihood parameter generation (MLPG) \cite{mlpg} algorithm is applied to our models to generate continuous curves from discrete sequences. In this way, smoother parameters can be generated to ensure the continuity of facial animation. To increase the variations, we use global variance of training data instead of predicted variance for MLPG algorithm. This is because we find that using global variance produces the same results as predicted variance but with less training time.

Another solution we have tried to eliminate discontinuities is sliding window regression method mentioned in \cite{disney}. Animaiton parameter sequences of overlapping fixed-length will be firstly predicted. After that, averaging frame-wise will be applied to make the sequence smoother. In our experiments, this method achieves same performance as MLPG when output window size is set to 15. However, sliding window regression method takes much time to train. So we finally choose MLPG for postprocessing.

After postprocessing, we apply the method as described in \cite{cons1}\cite{cons2}\cite{wu2019mvf} to the face warehouse parameters predicted by our model. As a result, a synchronized 3D facial animation is generated.

\section{Experiments and Analysis}

\subsection{Experiment Setup}

For SI-ASR based PPG extractor, we train our model on an ASR dataset including Mandarin and English speech recordings from more than 1000 speakers, which corresponds to a total of about 11000 hours. Universal phoneme space contains 179 Mandarin units and 39 English units.

For BLSTM based animation parameters predictor, we train the model on a Mandarin dataset including synchronized speech and face warehouse parameters consisting of 11000 sentences from single speaker corresponding to a total of about 18 hours. Besides, our dataset have 4 emotion labels: neutral, angry, happy, sad with the ratio of 2.5:1:1:1.

\subsection{Objective Experiments}

\subsubsection{Speaker Independency}

To validate the speaker independency, we conduct experiments to compare the mean square error (MSE) loss of proposed method and that of speaker-dependent method. MFCC is the most commonly used feature in speaker-dependent method, so we implement MFCC-BLSTM as baseline for comparison. MFCC-BLSTM has the same architecture as the proposed method except that the PPG is replaced with MFCC extracted from input speech. To evaluate two systems, we choose 500 samples for normal case (the speaker of the input speech is the same as the speaker of the training set) and unseen speaker respectively. Results are presented in Fig.\ref{fig:MSE}. From the results we can see MFCC-BLSTM performs better in normal case but get worse in unseen speaker case. That is reasonable since MFCC is speaker-dependent. And it demonstrates the advantage of proposed method on speaker independent cases. Loss increases rapidly in unseen speaker case since predicted parameters and groundtruth correspond to different 3D face respectively, and 3D face driven by inconsistent parameters will generate reasonable animations but different from original videos.

\subsubsection{Language Independency}

\begin{figure}[t]
  \centering
  \includegraphics[width=7cm, height=4.5cm]{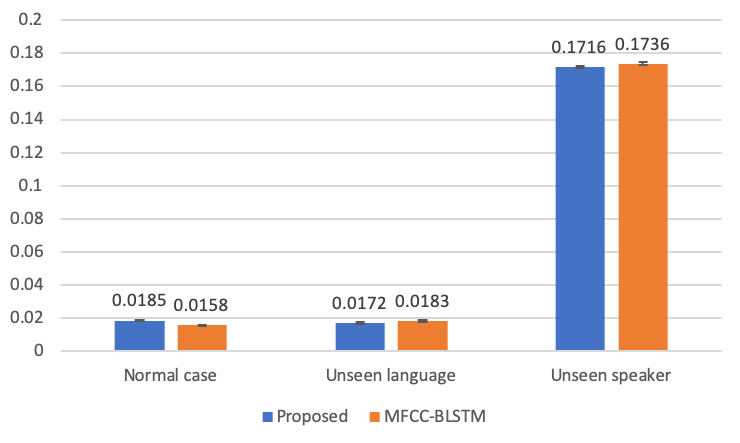}
  \vspace{5pt}
  \caption{MSE loss of two methods. Normal case means the speaker and language are the same as training set. Proposed method achieves better performance in unseen language or unseen speaker cases.}
  \vspace{-10pt}
  \label{fig:MSE}
\end{figure}

Another experiment is to validate language independency of the proposed method. For the same reason, we choose MFCC-BLSTM as baseline. Fig.\ref{fig:MSE} presents the results of two methods on 500 multilingual/mixlingual samples. From the results, proposed method has superior performance over MFCC-BLSTM under unseen language case. It indicates that our model is more robust to unseen languages due to the use of universal phoneme space. 

\subsubsection{Noise Robustness}

\begin{figure}[t]
  \centering
  \includegraphics[width=6cm, height=4.5cm]{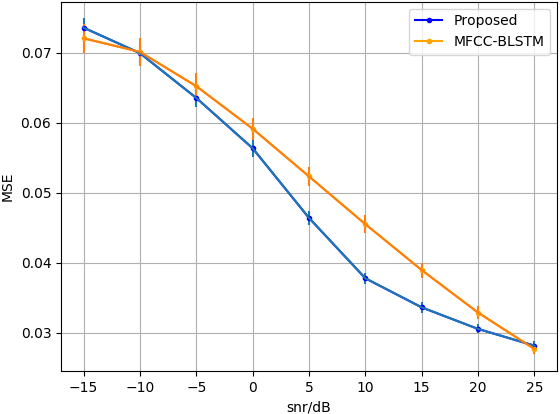}
  \vspace{10pt}
  \caption{MSE loss of two methods under different SNR conditions. When SNR is in the range of 10dB to 25dB, MSE loss of proposed method increases slower as SNR decreases. It shows that proposed method is more robust to noise at a high SNR level.}
  \label{fig:snr}
  \vspace{-15pt}
\end{figure}

To validate the noise robustness, we compare the performance of the proposed method and MFCC-BLSTM under noisy conditions. For evaluation, 500 utterances are randomly choosed from test set at first. Then we add noise to each utterance corresponding to a group of signal-to-noise ratio (SNR) from 25dB to -15dB (5dB step). After that each utterance will be mixed with 4 types of noise randomly, which includes multiple situations such as home, office and music. Preprocessing above results in a total of 2000 utterances for each SNR. Finally we compute the MSE Loss and results are presented in Fig.\ref{fig:snr}. We can see proposed method has significant improvement over MFCC-BLSTM under most cases. Especailly when SNR is controlled at a high level (from 25dB to 10dB in Fig.\ref{fig:snr}), loss of proposed method increases slower than MFCC as noise increases. It demonstrates that proposed method is more robust to noise at a normal level. As SNR continues to decrease, noise exceeds the ability of both features so it is reasonable that both methods achieve a similar performance at low level SNR (from -15dB to -5dB).

\subsection{Subjective Experiments}

We conduct mean opinion score (MOS) listening tests to validate the effectiveness of the proposed methods$\footnote{Samples can be found on https://thuhcsi.github.io/interspeech2020-talking-head-samples/ which cover all the following experiments}$. MFCC-BLSTM is still used as our baseline. Besides, we add groundtruth for reference. For each method, 16 videos including 4 emotions are evaluated. In MOS listening tests, 10 native Chinese speakers were asked to evaluate generated videos on a scale from 1 (Completely not expressive) to 5 (Completely expressive). To give a score, subjects need to focus on the performance of mouth shape, expressions and head pose. 

\subsubsection{Speaker Independency}

\begin{figure}[t]
  \centering
  \includegraphics[width=7cm, height=4.5cm]{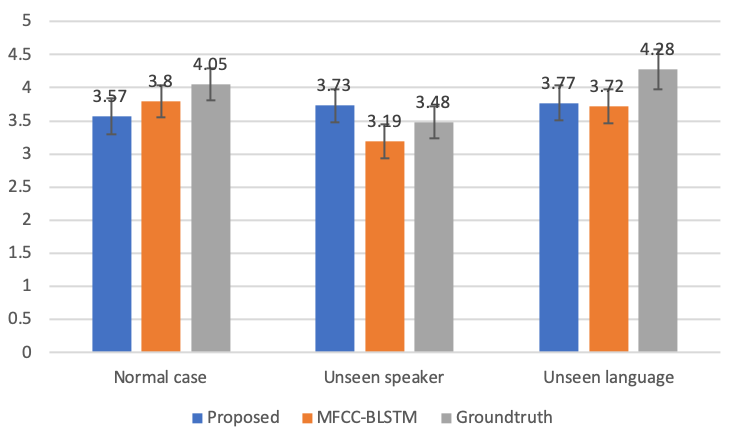}
  \vspace{3pt}
  \caption{MOS of two methods and groundtruth. Proposed method achieves better MOS than MFCC-BLSTM in unseen language or unseen speaker cases.}
  \label{fig:MOS}
  \vspace{-10pt}
\end{figure}

Results related to speaker independency are presented in Fig.\ref{fig:MOS}. Videos in unseen speakers include one unseen male and one unseen female speaker to cover different cases. Obviously proposed method outperforms MFCC-BLSTM under unseen speaker case. This conclusion validates the speaker-independency of PPG. Groundtruth's result decreases rapidly under unseen speaker case because we generate groundtruth videos using original speaker's talking head and unseen speaker's animation parameters. The inconsistency between 3D talking head and animation parameters results in bad performance.

\subsubsection{Language Independency}

To validate language independency, we generate videos from unseen language including multilingual and mixlingual cases. Limited by the dataset, experiments of unseen language are only conducted on English. As we can see, proposed method achieves a better performance under unseen language case. MFCC-BLSTM also receives a comparable result because phonemes from 2 languages have some similarities. In paticular, comparing unseen language (English) with normal case (Mandarin) we can see English samples receive higher MOS than Mandarin in most cases. The reason is that our subjects were all Chinese which resulted in stricter evaluation of Mandarin. The results indicate that universal phoneme space works under unseen languages.

\subsubsection{Energy Efficiency}

\begin{figure}[t]
  \centering
  \includegraphics[width=6.4cm, height=0.8cm]{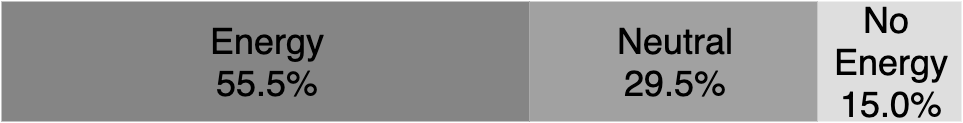}
  \vspace{5pt}
  \caption{Percentage preference of proposed PPG-BLSTM with/without energy. Compared to PPG-BLSTM without energy, PPG-BLSTM with energy effectively solves the problem that the mouth cannot keep closed during silent part of speech.}
  \label{fig:ABX}
  \vspace{-17pt}
\end{figure}

We also conduct an A/B preference test to validate the effectiveness of energy in silent part as mentioned in Section.2.2. We set 4 videos generated from PPG-BLSTM with/without energy for evaluation. Subjects are required to judge whether mouth is closed correctly in silent part. Two videos will be shuffled in each group and subjects should choose a preference or neutral (No preference) according to the mouth movement. Results are depicted in Fig.\ref{fig:ABX}, the model with energy has better performance, illustrating the effectiveness of energy.

\section{Conclusions}
In this paper, we propose a robust method which supports multilingual/mixlingual, speaker-independent speech-driven talking head generation with different emotions. Our method achieves speaker-independency by training an SI-ASR based PPG extractor with multi-speaker ASR dataset to extract PPG from speech as input features. By using multilingual ASR dataset and building universal phoneme space, our model also performs well under multilingual/mixlingual cases. Objective and subjective experiments show that our method can generate high-quality expressive animations under unseen speakers and languages. In the future work, we intend to explore more input features and model architectures to imporve the performance.

\section{Acknowledgements}
This work is supported by joint research fund of National Natural Science Foundation of China - Research Grant Council of Hong Kong (NSFC-RGC) (61531166002, N\_CUHK404/15), National Natural Science Foundation of China (61433018, 61375027). We would also like to thank Tencent AI Lab Rhino-Bird Focused Research Program (No. JR201942) and Tsinghua University - Tencent Joint Laboratory for the support.

\bibliographystyle{IEEEtran}
\bibliography{myrefs}

\begin{thebibliography}{10}
\providecommand{\url}[1]{#1}
\csname url@samestyle\endcsname
\providecommand{\newblock}{\relax}
\providecommand{\bibinfo}[2]{#2}
\providecommand{\BIBentrySTDinterwordspacing}{\spaceskip=0pt\relax}
\providecommand{\BIBentryALTinterwordstretchfactor}{4}
\providecommand{\BIBentryALTinterwordspacing}{\spaceskip=\fontdimen2\font plus
\BIBentryALTinterwordstretchfactor\fontdimen3\font minus
  \fontdimen4\font\relax}
\providecommand{\BIBforeignlanguage}[2]{{%
\expandafter\ifx\csname l@#1\endcsname\relax
\typeout{** WARNING: IEEEtran.bst: No hyphenation pattern has been}%
\typeout{** loaded for the language `#1'. Using the pattern for}%
\typeout{** the default language instead.}%
\else
\language=\csname l@#1\endcsname
\fi
#2}}
\providecommand{\BIBdecl}{\relax}
\BIBdecl

\bibitem{mattheyses2013comprehensive}
W.~Mattheyses, L.~Latacz, and W.~Verhelst, ``Comprehensive many-to-many
  phoneme-to-viseme mapping and its application for concatenative visual speech
  synthesis,'' \emph{Speech Communication}, vol.~55, no. 7-8, pp. 857--876,
  2013.

\bibitem{theobald2012relating}
B.-J. Theobald and I.~Matthews, ``Relating objective and subjective performance
  measures for aam-based visual speech synthesis,'' \emph{IEEE Transactions on
  Audio, Speech, and Language Processing (TASLP)}, vol.~20, no.~8, pp.
  2378--2387, 2012.

\bibitem{wu2019mvf}
F.~Wu, L.~Bao, Y.~Chen, Y.~Ling, Y.~Song, S.~Li, K.~N. Ngan, and W.~Liu,
  ``Mvf-net: Multi-view 3d face morphable model regression,'' in \emph{IEEE
  Conference on Computer Vision and Pattern Recognition (CVPR)}, 2019, pp.
  959--968.

\bibitem{biasutto2019modeling}
T.~Biasutto, S.~Dahmani, S.~Ouni \emph{et~al.}, ``Modeling labial
  coarticulation with bidirectional gated recurrent networks and transfer
  learning,'' in \emph{Annual Conference of International Speech Communication
  Association (Interspeech)}, 2019, pp. 2608--2612.

\bibitem{xu2013practical}
Y.~Xu, A.~W. Feng, S.~Marsella, and A.~Shapiro, ``A practical and configurable
  lip sync method for games,'' in \emph{Proceedings of Motion on Games}, 2013,
  pp. 131--140.

\bibitem{taylor2012dynamic}
S.~L. Taylor, M.~Mahler, B.-J. Theobald, and I.~Matthews, ``Dynamic units of
  visual speech,'' in \emph{Proceedings of the 11th ACM SIGGRAPH/Eurographics
  conference on Computer Animation}, 2012, pp. 275--284.

\bibitem{schabus2011simultaneous}
D.~Schabus, M.~Pucher, and G.~Hofer, ``Simultaneous speech and animation
  synthesis,'' in \emph{ACM Annual Conference Special Interest Group on
  Computer Graphics and Interactive Techniques (SIGGRAPH)}, 2011, pp. 1--1.

\bibitem{wang2012high}
L.~Wang, W.~Han, and F.~K. Soong, ``High quality lip-sync animation for 3d
  photo-realistic talking head,'' in \emph{IEEE International Conference on
  Acoustics, Speech and Signal Processing (ICASSP)}, 2012, pp. 4529--4532.

\bibitem{merritt2013investigating}
T.~Merritt and S.~King, ``Investigating the shortcomings of {HMM} synthesis,''
  in \emph{Eighth ISCA Workshop on Speech Synthesis}, 2013, pp. 165--170.

\bibitem{dahmani2019conditional}
S.~Dahmani, V.~Colotte, V.~Girard, and S.~Ouni, ``Conditional variational
  auto-encoder for text-driven expressive audiovisual speech synthesis,'' in
  \emph{Annual Conference of International Speech Communication Association
  (Interspeech)}, 2019, pp. 2598--2602.

\bibitem{obamanet}
R.~Kumar, J.~Sotelo, K.~Kumar, A.~de~Br{\'{e}}bisson, and Y.~Bengio,
  ``Obamanet: Photo-realistic lip-sync from text,'' \emph{CoRR}, vol.
  abs/1801.01442, 2018.

\bibitem{obamalip}
S.~Suwajanakorn, S.~M. Seitz, and I.~Kemelmacher-Shlizerman, ``Synthesizing
  obama: learning lip sync from audio,'' \emph{ACM Transactions on Graphics
  (TOG)}, vol.~36, no.~4, pp. 1--13, 2017.

\bibitem{jali}
P.~Edwards, C.~Landreth, E.~Fiume, and K.~Singh, ``Jali: an animator-centric
  viseme model for expressive lip synchronization,'' \emph{ACM Transactions on
  Graphics (TOG)}, vol.~35, no.~4, pp. 1--11, 2016.

\bibitem{fehp2018}
D.~Greenwood, I.~Matthews, and S.~Laycock, ``Joint learning of facial
  expression and head pose from speech,'' in \emph{Annual Conference of
  International Speech Communication Association (Interspeech)}, 2018, pp.
  2484--2488.

\bibitem{disney}
S.~Taylor, T.~Kim, Y.~Yue, M.~Mahler, J.~Krahe, A.~G. Rodriguez, J.~Hodgins,
  and I.~Matthews, ``A deep learning approach for generalized speech
  animation,'' \emph{ACM Transactions on Graphics (TOG)}, vol.~36, no.~4, pp.
  1--11, 2017.

\bibitem{xielei2016}
B.~Fan, L.~Xie, S.~Yang, L.~Wang, and F.~K. Soong, ``A deep bidirectional
  {LSTM} approach for video-realistic talking head,'' \emph{Multimedia Tools
  and Applications (MTA)}, vol.~75, no.~9, pp. 5287--5309, 2016.

\bibitem{yousay}
J.~S. Chung, A.~Jamaludin, and A.~Zisserman, ``You said that?'' in
  \emph{British Machine Vision Conference}, 2017.

\bibitem{vgg}
K.~Chatfield, K.~Simonyan, A.~Vedaldi, and A.~Zisserman, ``Return of the devil
  in the details: Delving deep into convolutional nets,'' in \emph{British
  Machine Vision Conference}, 2014.

\bibitem{ppg}
L.~Sun, K.~Li, H.~Wang, S.~Kang, and H.~Meng, ``Phonetic posteriorgrams for
  many-to-one voice conversion without parallel data training,'' in \emph{IEEE
  International Conference on Multimedia and Expo (ICME)}, 2016, pp. 1--6.

\bibitem{luhui}
H.~Lu, Z.~Wu, R.~Li, S.~Kang, J.~Jia, and H.~Meng, ``A compact framework for
  voice conversion using wavenet conditioned on phonetic posteriorgrams,'' in
  \emph{IEEE International Conference on Acoustics, Speech and Signal
  Processing (ICASSP)}, 2019, pp. 6810--6814.

\bibitem{mlpg}
K.~Tokuda, T.~Yoshimura, T.~Masuko, T.~Kobayashi, and T.~Kitamura, ``Speech
  parameter generation algorithms for {HMM}-based speech synthesis,'' in
  \emph{IEEE International Conference on Acoustics, Speech, and Signal
  Processing. Proceedings (ICASSP)}, 2000, pp. 1315--1318.

\bibitem{zone}
D.~Krueger, T.~Maharaj, J.~Kram{\'{a}}r, M.~Pezeshki, N.~Ballas, N.~R. Ke,
  A.~Goyal, Y.~Bengio, A.~C. Courville, and C.~J. Pal, ``Zoneout: Regularizing
  rnns by randomly preserving hidden activations,'' in \emph{International
  Conference on Learning Representations (ICLR)}, 2017.

\bibitem{fwh32}
C.~Cao, Y.~Weng, S.~Zhou, Y.~Tong, and K.~Zhou, ``Facewarehouse: A {3D} facial
  expression database for visual computing,'' \emph{IEEE Transactions on
  Visualization and Computer Graphics (TVCG)}, vol.~20, no.~3, pp. 413--425,
  2013.

\bibitem{cons1}
H.~Kim, P.~Garrido, A.~Tewari, W.~Xu, J.~Thies, M.~Nie{\ss}ner, P.~P{\'e}rez,
  C.~Richardt, M.~Zollh{\"o}fer, and C.~Theobalt, ``Deep video portraits,''
  \emph{ACM Transactions on Graphics (TOG)}, vol.~37, no.~4, pp. 1--14, 2018.

\bibitem{cons2}
O.~Fried, A.~Tewari, M.~Zollh{\"o}fer, A.~Finkelstein, E.~Shechtman, D.~B.
  Goldman, K.~Genova, Z.~Jin, C.~Theobalt, and M.~Agrawala, ``Text-based
  editing of talking-head video,'' \emph{ACM Transactions on Graphics (TOG)},
  vol.~38, no.~4, pp. 1--14, 2019.

\end{thebibliography}

\end{document}